\documentclass[floatfix,superscriptaddress,twocolumn,aps,pre]{revtex4-2}

\pdfoutput=1

\usepackage{graphicx}
\usepackage{bm}
\usepackage{physics}
\usepackage[mathlines]{lineno}
\usepackage{setspace}
\usepackage[utf8]{inputenc}
\usepackage[english]{babel}
\usepackage[T1]{fontenc}
\usepackage[dvipsnames]{xcolor}
\usepackage{amssymb}
\usepackage{amsmath}

\usepackage{textalpha}
\newcommand{\mean}[1]{\left\langle #1 \right\rangle}

\newcommand{\ra}{\rightarrow}
\NewDocumentCommand{\dsum}{%
    e{^_}
}{%
  {%
    \displaystyle\sum
    \IfValueT{#1}{^{#1}}
    \IfValueT{#2}{_{#2}}
  }
}%

\begin{document}

\title{Quantifying ionization in hot dense plasmas}

\author{Thomas Gawne}
\email{thomas.gawne@physics.ox.ac.uk}
\affiliation{Department of Physics, Clarendon Laboratory, University of Oxford, Parks Road, Oxford OX1 3PU, UK}

\author{Sam M. Vinko}
\affiliation{Department of Physics, Clarendon Laboratory, University of Oxford, Parks Road, Oxford OX1 3PU, UK}
\affiliation{Central Laser Facility, STFC Rutherford Appleton Laboratory, Didcot OX11 0QX, UK}

\author{Justin S. Wark}
\affiliation{Department of Physics, Clarendon Laboratory, University of Oxford, Parks Road, Oxford OX1 3PU, UK}

\date{\today}

\begin{abstract}

Ionization is a problematic quantity in that it does not have a well-defined thermodynamic definition, yet it is a key parameter within plasma modelling. One still therefore aims to find a consistent and unambiguous definition for the ionization state. Within this context we present finite-temperature density functional theory calculations of the ionization state of carbon in CH plasmas using two potential definitions: one based on counting the number of continuum electrons, and another based on the optical conductivity.
Differences of up to 10\% are observed between the two methods. However, including ``Pauli forbidden'' transitions in the conductivity reproduces the counting definition, suggesting such transitions are important to evaluate the ionization state.

\end{abstract}

\maketitle

\section{\label{sec:intro}Introduction}

Dense plasmas comprise complex, inherently quantum states of matter, covering a wide variety of different temperatures and densities. An exact treatment of a plasma would involve solving a many-body Schr{\"o}dinger or Dirac equation that includes the full interactions between the particles. Given the extraordinarily large number of electrons and ions in realistic plasmas, this is impossible.
In lieu of directly solving the many-body equation, simpler models that only deal with the relevant physics required to make a set of predictions are used.
Well-known examples include finite-temperature density functional theory (FT-DFT)~\cite{HohenbergKohn,KohnSham}, collisional-radiative atomic-kinetics models~\cite{SCFLY}, and hydrodynamic modelling~\cite{marinak1998comparison}.
Of course, it is important to be sure that plasma models are able to make good predictions.
This is important for the general understanding of plasmas~\cite{toleikis2010probing,ciricosta2016detailed,hollebon2019ab}, and is especially pertinent in light of recent successes towards ignition in inertial confinement fusion~\cite{abu2022lawson,zylstra2022experimental}, where strong plasma modelling continues to be critical to improving gain~\cite{kritcher2022design}.

Within plasma modelling there are a number of parameters that are used to describe a plasma and make predictions about experimental observables; often the temperature $T$, density $\rho$, and the ionization state of the ions $Z$. However, unlike the first two parameters, $Z$ does not have a well-defined thermodynamic definition~\cite{griem1997principles}. This is not a trivial issue. The ionization state is an input parameter to many equations, so the choice of definition will have a cascading effect on the evolution of a plasma. It is also important to understand how the choice of definition used in plasma modelling relates to experimental observations.
As plasma models deal with time-dependent ionization dynamics, they must have an unambiguous definition of ionization. To that end, the ionization state is often defined by the number of electrons bound to ions, with the remaining electrons considered purely free. The number of electrons free from an ion is its charge state (i.e. $Z=0$ for a neutral atom), and the mean charge state of the system is used to represent its mean ionization state (MIS).
This definition is then fed directly into equations governing the physical properties of the plasma. For example, in this bound-or-free electron picture, continuum lowering is thought to be treatable using models of ionization potential depression (IPD) such as the Stewart-Pyatt (SP) model~\cite{stewart1966lowering}.
However, in the past decade, a substantial body of research has emerged indicating a lack of consensus between widely-employed IPD models and experimental observations in dense plasma systems~\cite{ciricosta2013direct,hoarty2013observations,ciricosta2016measurements,kraus2016x,doppner2023observing}, raising concerns over our predictive capabilities in this challenging regime.

An experiment at the National Ignition Facility (NIF)~\cite{kraus2016x} found that the MIS of hot dense CH plasmas inferred from x-ray Thomson scattering (XRTS) measurements was substantially higher than predicted by plasma modelling. The discrepancy was attributed to problems with the IPD models used in evaluating the ionization, as simply increasing the MIS seemed to reproduce the XRTS data. However, more recent experimental measurements in hot dense Be~\cite{doppner2023observing} found that XRTS spectra could not be reproduced by artificially increasing the MIS, suggesting the reason for the discrepancies between these experiments and plasma models is more complex.

The approach of assuming bound-or-free electrons provides a conceptually simple and intuitive definition of ionization.
Therefore, it is often used in plasma models where the effect of delocalized electrons needs to be considered, such as in collisional-radiative atomic-kinetics simulations~\cite{FLYCHK}, IPD models, in the Chihara decomposition~\cite{chihara1987difference,chihara2000interaction} for XRTS modelling, and in atomic cascade calculations~\cite{fritzsche2021atomic}.
In the limit that electronic states can be distinguished as strongly localized and around ions or fully delocalized, the bound-or-free approach should be adequate to represent the MIS of a system. However, recent first principles calculations have shown that such a simple separation of electronic states is generally not possible in hot dense systems~\cite{gawne2023investigating}.
At the same time, it is not clear that the mean charge state should always be representative of the ionization state accessed in spectroscopy experiments involving high energy density systems. The motivation of this work is in exploring whether a definition of ionization based on bound or free electrons can be a suitable definition.

For the present investigation, the equivalent definition of the mean charge state in a first-principles DFT calculation is to count the number of electrons in the continuum ($c$) bands $N_{\rm cond}$:
\begin{equation}
    \label{eq:MIScount}
    \mean{Z}_{\rm count} = \frac{1}{N_{i}} \sum_{\bm k} w_{\bm k} \sum_{n \in c} f(\epsilon_{\bm{k}, n})  = \frac{N_{\rm cond}}{N_i} \, ,
\end{equation}
where $N_i$ is the number of ions, $\epsilon_{\bm{k}, n}$ is the eigenenergy of the Bloch state $\ket{\psi_{\bm{k}, n}}$, $n$ is the band number, $\bm{k}$ is the crystal momentum, $w_{\bm k}$ is the $k$-point weight, and $f(\epsilon_{\bm{k}, n})$ is the Fermi-Dirac occupation number of the state and includes the state's degeneracy.

Recently, attempts have been made to define ionization based on other physical properties of systems~\cite{bethkenhagen2020carbon, clerouin2022equivalence}. One such definition~\cite{bethkenhagen2020carbon} is based on the real component of the frequency-dependent optical conductivity $\sigma(\omega)$. As the conductivity can be measured experimentally, it is proposed that this definition, outlined below, would provide direct access to the MIS. To derive the MIS, the optical conductivity is assumed to be described by the Kubo-Greenwood (KG) formula~\cite{kubo1957statistical,greenwood1958boltzmann}:
\begin{equation}
    \label{eq:KG_Cond}
    \begin{aligned}
    \sigma(\omega) &= \frac{\pi \hbar e^2}{m_e V} \sum_{\bm k} w_{\bm k}  \sum_{n,m} \left[f(\epsilon_{\bm{k}, m}) - f(\epsilon_{\bm{k}, n}) \right] g^{\bm k}_{nm} \\
    &\times \delta\left(\epsilon_{\bm{k}, n} - \epsilon_{\bm{k}, m} - \hbar \omega \right) \, ,
    \end{aligned}
\end{equation}
where $V$ is the volume of the system, and $g^{\bm k}_{nm}$ are the dipole transition matrix elements:
\begin{equation}
    \label{eq:MatrixElements}
    g^{\bm k}_{nm} = \frac{\hbar^2}{3m_e} \frac{\left| \bra{\psi_{\bm{k}, m}} \nabla \ket{\psi_{\bm{k}, n}} \right|^2}{\epsilon_{\bm{k}, n} - \epsilon_{\bm{k}, m}} \, .
\end{equation}
For Bloch states $\ket{\psi_{\bm{k}, n}} = e^{i \bm{k} \cdot \bm{r}} \ket{u_{\bm{k}, n}}$, it can be shown that~\cite{blount1962formalisms,ashcroft1976solid}:
\begin{equation}
    \label{eq:BlochVelocity}
    \begin{aligned}
    \frac{\hbar^2}{i m_e} \bra{\psi_{\bm{k}, m}} \nabla \ket{\psi_{\bm{k}, n}} &= \delta_{m,n} \nabla_{\bm k} \epsilon_{\bm{k}, m} \\
    &+ (\epsilon_{\bm{k}, n} - \epsilon_{\bm{k}, m}) \braket{u_{\bm{k}, m}}{\nabla_{\bm k} u_{\bm{k}, n}}.
    \end{aligned}
\end{equation}
The Thomas-Reiche-Kuhn (TRK) sum rule~\cite{thomas1925zahl, reiche1925zahl, kuhn1925gesamtstarke} states that, for a complete basis set, the sum of the dipole matrix elements is unity. In momentum form, this is:
\begin{equation}
    \label{eq:TRK}
    2 \sum_{\substack{n \\ \epsilon_{\bm{k}, n} \neq \epsilon_{\bm{k}, m}}} g^{\bm k }_{nm} 
    = 2 \sum_{n} g^{\bm k }_{nm} \tau(\epsilon_{\bm{k}, n} - \epsilon_{\bm{k}, m}) = 1 \, ,
\end{equation}
where $\tau(\epsilon_{\bm{k}, n} - \epsilon_{\bm{k}, m}) = 1 - \delta(\epsilon_{\bm{k}, n} - \epsilon_{\bm{k}, m})$ ensures the $\epsilon_{\bm{k}, n} = \epsilon_{\bm{k}, m}$ terms are excluded.
The number of electrons in the system $N_e$ can therefore be recovered by including a sum over the occupation numbers:
\begin{equation}
    \label{eq:TRK_electron_num}
    \begin{aligned}
     N_e &= \sum_{\bm k} w_{\bm k}  \sum_{m} f(\epsilon_{\bm{k}, m}) \\
     &= 2 \sum_{\bm k} w_{\bm k} \sum_{m} f(\epsilon_{\bm{k}, m}) \sum_{n} g^{\bm k }_{nm} \tau(\epsilon_{\bm{k}, n} - \epsilon_{\bm{k}, m}) \\
     &= \sum_{\bm k} w_{\bm k} \sum_{n,m} \left[f(\epsilon_{\bm{k}, n}) - f(\epsilon_{\bm{k}, m}) \right] g^{\bm k}_{mn} \tau(\epsilon_{\bm{k}, n} - \epsilon_{\bm{k}, m}) \, ,
     \end{aligned}
\end{equation}
where the relationship $g^{\bm k}_{nm}=-g^{\bm k}_{mn}$ for $n \neq m$ and index swapping are used to derive the last line.
A similar-looking sum can be constructed by integrating the KG conductivity over all frequencies:
\begin{equation}
    \begin{aligned}
    S &= \frac{2 m_e V}{\pi e^2} \int_{0}^{\infty} \sigma(\omega) d\omega \\
     &=\sum_{\bm k} w_{\bm k}  \sum_{n,m} \left[f(\epsilon_{\bm{k}, m}) - f(\epsilon_{\bm{k}, n}) \right] g^{\bm k}_{nm} \, ,
    \end{aligned}
\end{equation}
where the delta function in Eq.~(\ref{eq:KG_Cond}) is used to derive the second line.
Ref.~\cite{bethkenhagen2020carbon} has proposed that by splitting the conductivity into transitions between conduction ($c$) and valence ($v$) states, so that $\sigma = \sigma^{c \ra c} + \sigma^{v \ra c} + \sigma^{v \ra v}$, the MIS can be found by only considering the $c \ra c$ conductivity:
\begin{equation}
    \label{eq:MIStrk}
    \mean{Z}_{\rm TRK} = \frac{2 m_e V}{\pi e^2 N_i} \int_{0}^{\infty} \sigma^{c \ra c}(\omega) d\omega
    = \frac{N_{\rm eff}}{N_i}  \, ,
\end{equation}
where $N_{\rm eff}$ is the number of conduction electrons calculated in this scheme.
Note that this is akin to deciding which states should be considered bound or free.
However, care is needed: the sums in $S$ include the $\epsilon_{\bm{k}, n} = \epsilon_{\bm{k}, m}$ terms, whereas the TRK sum rule excludes them.
When $n=m$, Eq.~(\ref{eq:BlochVelocity}) shows that $\bra{\psi_{\bm{k}, n}} \nabla \ket{\psi_{\bm{k}, n}}$ is not always zero, but is proportional to the curvature of the band at $\bm{k}$.
Meanwhile, in the limit of zero energy difference, the difference in the eigenenergies in $g^{\bm k}_{nn}$ and the occupation numbers in Eq.~(\ref{eq:KG_Cond}) resolve to the gradient of the Fermi-Dirac distribution at $\epsilon_{\bm{k},n}$, so the $n=m$ terms are finite and not necessarily zero~\cite{calderin2017kubo}. Therefore, if the $g^{\bm k}_{nn}$ terms are included, $S \ge N_e$~\cite{calderin2017kubo}.
In practice, converging the sums $N_e$ and $S$ requires a huge number of bands, even at moderate temperatures, so it can appear that $S < N_e$. If the $n=m$ terms are included, the number of electrons extracted from $\sigma^{c \ra c}$ (which includes $g^{\bm k}_{nn}$ terms) may appear larger than it should when compared with the equivalent TRK sum.

The conductivity-based definition in Eq.~(\ref{eq:MIStrk}) has recently been applied to predict the MIS found via XRTS measurements of hot dense Be plasmas generated in implosions at the NIF~\cite{doppner2023observing}.
The conductivity measure was found to agree well with the XRTS-inferred MIS at the highest temperatures and densities produced ($T = 150$~eV, $\rho \ge 30$~g~cm$^{-3}$), while the counting measure using the SP model under-predicts the MIS. However, at lower temperatures and densities ($T = 100$~eV, $\rho \sim 10$~g~cm$^{-3}$) the counting measure agrees well with the experimental data, while the conductivity definition appears to over-predict the MIS.
Furthermore, at mass densities $\rho \lesssim 10$~g~cm$^{-3}$, differences in the predictions between the counting and conductivity definitions grow further as the temperature decreases, even in conditions where the counting definition is expected to be applicable~\cite{doppner2023observing}.

Calculating the effective ionization state of a system using the $c \ra c$ conductivity has been explored previously in condensed matter physics. Notably, when applied to experimental measurements of the conductivity in alkali metals in the early 70's, it was found that $N_{\rm eff} > 1$ \cite{whang1972optical}, causing consternation since this implies that the ionization state exceeds the number of electrons in the conduction band, $N_{\rm cond}$. Kobayasi and Watabe~\cite{kobayasi1973remark} showed how this problem could be resolved by completing the TRK sum by including Pauli forbidden transitions in the calculation of the ionization state. An equivalent term does not appear to have been included in the ionization calculations of Refs.~\cite{bethkenhagen2020carbon,doppner2023observing}.

In this letter, we calculate the MIS of hot dense CH plasmas from first-principles using FT-DFT, using both the counting definition in Eq.~(\ref{eq:MIScount}) and the conduction-based definition of Eq.~(\ref{eq:MIStrk}). Differences of up to 10\% are seen in the MIS between these two methods, with the differences becoming larger at low temperatures. We extend Kobayasi and Watabe's arguments to finite temperatures, and examine the effect on the ionization state predicted by the conductivity. Like Kobayasi and Watabe, we find that including $v \rightarrow c$ transitions recovers the counting result.

\section{Counting the Conduction Electrons}

To calculate the effective number of conduction electrons, the TRK sum over states is split into a sum over conduction states ($n \in c$) and valence states ($n \in v$):
\begin{equation}
    \begin{aligned}
    N_e &= \sum_{\bm k} w_{\bm k} \sum_{n,m} \left[f(\epsilon_{\bm{k}, n}) - f(\epsilon_{\bm{k}, m}) \right] g^{\bm k}_{mn} \tau(\epsilon_{\bm{k}, n} - \epsilon_{\bm{k}, m}) \\
      &= \sum_{\bm k} w_{\bm k} \left( \sum_{n \in c} + \sum_{n \in v} \right)\left( \sum_{m \in c} + \sum_{m \in v} \right) g^{\bm k}_{mn} \\
      &\quad \times \left[f(\epsilon_{\bm{k}, n}) - f(\epsilon_{\bm{k}, m}) \right] \tau(\epsilon_{\bm{k}, n} - \epsilon_{\bm{k}, m}) \, ,
    \end{aligned}
\end{equation}
It is proposed that only the $c \ra c$ transitions need to be used to calculate the number of conduction electrons~\cite{kobayasi1973remark,bethkenhagen2020carbon}:
\begin{equation}
    \label{eq:Neff}
    \begin{aligned}
    N_{\rm eff} &= \sum_{\bm k} w_{\bm k} \sum_{n \in c} \sum_{m \in c} \left[f(\epsilon_{\bm{k}, n}) - f(\epsilon_{\bm{k}, m}) \right] g^{\bm k}_{mn} \\
    &\quad \times \tau(\epsilon_{\bm{k}, n} - \epsilon_{\bm{k}, m}) \\ 
    &= 2 \sum_{\bm k} w_{\bm k} \sum_{n \in c} f(\epsilon_{\bm{k}, n}) \sum_{m \in c} g^{\bm k}_{mn} \tau(\epsilon_{\bm{k}, n} - \epsilon_{\bm{k}, m}) \, .
    \end{aligned}
\end{equation}
However, Eq.~(\ref{eq:Neff}) does not include a complete sum rule~\cite{kobayasi1973remark}. To do so, the valence states need to be included again using:
\begin{equation}
    \begin{aligned}
    \label{eq:Neff_corr}
    N_{\rm eff}  &= 2 \sum_{\bm k} w_{\bm k} \sum_{n \in c} f(\epsilon_{\bm{k}, n}) \left( \sum_{m} - \sum_{m \in v} \right) g^{\bm k}_{mn} \\
    &\quad \times \tau(\epsilon_{\bm{k}, n} - \epsilon_{\bm{k}, m}) \\ 
    &= \sum_{\bm k} w_{\bm k} \sum_{n \in c} f(\epsilon_{\bm{k}, n}) \\
    &\quad - 2 \sum_{\bm k} w_{\bm k} \sum_{n \in c} f(\epsilon_{\bm{k}, n}) \sum_{m \in v} g^{\bm k}_{mn} \tau(\epsilon_{\bm{k}, n} - \epsilon_{\bm{k}, m}) \\
    &\equiv N_{\rm cond} - \Delta N_{\rm eff} \, .
    \end{aligned}
\end{equation}
The $\Delta N_{\rm eff}$ term relates the number of ionized electrons from the conduction definition to the counting definition. $\Delta N_{\rm eff} < 0$, so it represents an increase in the  number of conduction electrons from the optical conductivity as compared with just counting them.
In the cold limit, this term only includes transitions between the valence states (which are fully occupied) and the occupied conduction states. In other words, this term only includes forbidden transitions due to Pauli blocking~\cite{kobayasi1973remark}.
In the finite-temperature limit, this term does not quite represent transitions between Pauli blocked states as there is no statistical weighting accounting for the thermal de-occupation of the valence states at high temperatures.
This derivation reveals the link between the conductivity-based definition of ionization proposed in Ref.~\cite{bethkenhagen2020carbon} and the simple electron counting method.

\section{Results and Discussion}

\begin{figure}
    \centering
    \includegraphics[width=\linewidth,keepaspectratio]{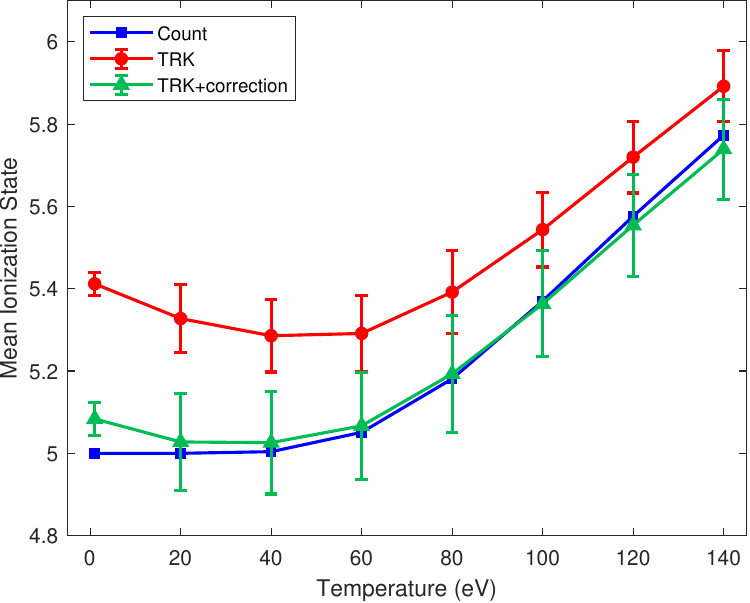}
    \caption{Mean ionization state of CH ($N/N_{\rm CH}$) calculated using different methods. The red circles represent the MIS using $N = N_{\rm eff}$ as suggested in Ref.~\cite{bethkenhagen2020carbon}. The green triangles use $N = N_{\rm eff} + \Delta N_{\rm eff}$. The blue squares denote the counting definition, $N = N_{\rm cond}$.}
    \label{fig:MIS}
\end{figure}

We proceed to examine the degree to which the $c \ra c$ conductivity over-predicts the ionization state via FT-DFT simulations of CH plasmas at different temperatures, performed using the \textsc{Abinit} v8.10.3 package~\cite{gonze2016recent,torrent2008implementation,bottin2008large}.
We choose CH so that we can compare our calculations with experimental measurements of the MIS from Ref.~\cite{kraus2016x}.
We simulate supercells containing 32 atoms (C$_{16}$H$_{16}$), with the lattice parameters chosen to give a mass density of 6.74~g~cm$^{-3}$. The electron temperature was varied from 1-140~eV. The density and temperatures were chosen to match the experimental conditions of Ref.~\cite{kraus2016x}. 
In order to have a well-converged TRK sum rule the vast majority of occupied bands need to be calculated explicitly.
For each calculation the number of bands used ensured the sum rule was at least 97\% complete (4500-7000 bands).
To ensure there are no high-level symmetries in the system the ions were randomly placed in the cell. Although this effects the specific shapes of the density of states and the conductivity, the MIS was found to be largely independent of the exact ion positions. The \textit{Atompaw} code~\cite{atompaw} was used to generate projector augmented-wave (PAW)~\cite{BlochlPAW} potentials for the C and H atoms.
Due to the high temperatures involved, all of the electrons are treated as valence. The PAW potentials were generated for atoms at zero-temperature, and previous studies have shown that such potentials can be applied to investigate thermal ionization effects~\cite{hollebon2021excited}.
A PAW core radius of 0.5 Bohr is used for both atoms. For the ion positions that were sampled, there was no overlap of the cores.
The PBE form of the generalized gradient approximation~\cite{PBE-GGA} was used for the exchange-correlation functional.

The mean ionization state of CH versus the electron temperature using the different definitions is plotted in Fig.~(\ref{fig:MIS}). To avoid over-counting the electrons when integrating the conductivity, the TRK definitions were calculated using the sums over states, as in Eqs.~(\ref{eq:Neff}) and (\ref{eq:Neff_corr}).
The error bars on the TRK calculations represent the uncertainty in the MIS due to the sum rule being incomplete. The red circles denote the MIS calculated using $N_{\rm eff}$, and the blue squares from the counting definition, Eq.~(\ref{eq:MIScount}). Across the entire temperature range the $c \ra c$ conductivity definition  predicts a higher MIS than the counting definition, up to 10\% at the lowest temperatures. This appears to be in agreement with previous observations~\cite{doppner2023observing, kobayasi1973remark, whang1972optical}.
Additionally, the red curve has an interesting behaviour whereby the MIS increases as the temperature decreases below 40~eV. This is in contrast to the blue curve which monotonically increases with temperature, and is approximately constant at $\mean{Z}_{\rm count} = 5.0$ below 40~eV.
When the correction term $\Delta N_{\rm eff}$ is included (green triangles), the MIS from the conductivity reproduces the counting definition. There is a small deviation still at 1~eV, though this is likely due to the sum rule still being incomplete.
The reproduction of the counting MIS should not be surprising as the $\Delta N_{\rm eff}$ term directly connects $N_{\rm eff}$ to $N_{\rm cond}$.
These results suggest that the $c \ra c$ conductivity-based definition, as proposed in Ref.~\cite{bethkenhagen2020carbon}, does not provide the correct ionization state as it does not use a complete sum rule. We note that the correction term $\Delta N_{\rm eff}$ can be substantial, even at relatively high temperatures.

We can compare our results with experimental measurements of the CH ionization state from Ref.~\cite{kraus2016x}. Their hydrodynamic simulations suggested a mean mass density of 6.74~g~cm$^{-3}$, at which their XRTS spectrum could be fitted with a mean temperature of $86 \pm 20$~eV and a MIS of $5.92 \pm 0.15$. This ionization state is substantially larger than is predicted by our calculations using the counting definition.
The MIS calculated from $N_{\rm eff}$ is closer to the experimental data, though it is is still lower across the inferred temperature range.
Once the correction $\Delta N_{\rm eff}$ is applied, the conductivity-derived ionization is further from the experimental data.
In Ref.~\cite{doppner2023observing}, the MIS from $N_{\rm eff}$ appears to agree well with the XRTS spectrum of Be at very high densities and temperatures.
We ascribe this apparent improvement in the conductivity definition over the electron counting definition in these extreme conditions precisely to the fact that $N_{\rm eff}$ overestimates the number of continuum electrons as compared with $N_{\rm count}$.
At lower temperatures it becomes apparent that $N_{\rm eff}$ overestimates the MIS. Including a term involving transitions between valence and occupied conduction states -- transitions which at low temperatures are Pauli forbidden~\cite{kobayasi1973remark} -- allows for the conductivity to reproduce the number of electrons in the continuum. Our results suggest these ``Pauli forbidden'' transitions are important in evaluating the MIS from the conductivity at finite temperatures.

Of course, there remains the question of what are the additional electrons that XRTS measurements seem to suggest there should be compared with the counting definition. Ref.~\cite{doppner2023observing} showed that simply increasing the MIS would still not reproduce their XRTS data as the elastic feature would need to be larger than was measured.
Instead, they found that the discrepancies in the XRTS spectrum can be explained by the delocalization of the K-shell wavefunctions, caused by the continuum electron density screening the atomic nuclei.
As the screening electron density increases with temperature and compression, so does the delocalization, resulting in greater deviations of the XRTS spectrum from modelling. 
Using a simple self-consistent screening model, Ref.~\cite{doppner2023observing} were able reproduce their XRTS spectra.
In IPD models, the bound electrons are considered to be strongly localized around atomic sites, hence they cannot account for their gradual delocalization.
Among other effects, the delocalization of the K-shell wavefunctions results in them overlapping, allowing their electrons to move between atomic sites.
For completeness, we note that electrons moving within the K-shell states would be represented by the $v \ra v$ transitions in the optical conductivity, which are not included in $N_{\rm eff}$. In the present calculations, the contribution of these transitions is extremely small, but non-zero at temperatures above 40~eV. So while these contributions are not large enough to meaningfully increase the MIS directly, they still indicate the K-shell electrons are mobile at high temperatures.
We would therefore agree that more detailed accounting for the effect of atomic electrons may explain the higher ionization seen in XRTS experiments compared with counting the continuum electrons.

To conclude, we have examined the applicability of the bound-or-free electron model to describing the MIS in high energy-density systems by considering two potential definitions of ionization: one based on counting the number of electrons in the continuum, and another recently proposed definition based on the optical conductivity~\cite{bethkenhagen2020carbon}. For the latter definition, it is shown that unless transitions between the valence and conduction states are included -- transitions that at low temperatures are Pauli forbidden~\cite{kobayasi1973remark} -- then this definition will over-count the number of electrons in the continuum. It is also shown that only counting the continuum electrons predicts a lower MIS compared to that inferred from XRTS measurements at high temperatures and densities. However, in such extreme conditions, electrons in states that may be typically classed as bound gradually delocalize~\cite{doppner2023observing}. In the calculations presented here, at sufficiently high temperatures, the valence-valence conductivity is non-zero, which implies the electrons in the valence bands are mobile. We therefore conclude that more detailed accounting of the contribution of atomic electrons to the MIS may explain the higher ionization states in XRTS experiments compared with counting the continuum electrons. While this work focuses on CH to allow us to link to experiments, the results are general and are readily applicable to other materials across the full range of plasma conditions.

\section*{Acknowledgements}
T.G., J.S.W. and S.M.V. acknowledge support from AWE via the Oxford Centre for High Energy Density Science (OxCHEDS).
S.M.V. acknowledges support from the Royal Society.
J.S.W. and S.M.V. acknowledge support from the UK EPSRC under grants EP/P015794/1 and EP/W010097/1.
S.M.V. is a Royal Society University Research Fellow.

\bibliographystyle{apsrev4-2}
\bibliography{refs.bib}

\end{document}